\documentclass{WileyMSP-template}
\usepackage{color}
\usepackage{graphicx}
\usepackage{cite}

\begin{document}

\pagestyle{fancy}

\rmfamily
\title{Broadband Terahertz Quantum Cascade Laser Dual-Comb Sources under Off-Resonant Microwave Injection}
\maketitle

\author{Xiaoyu Liao$^{+}$}
\author{Ziping Li$^{+}$}
\author{Kang Zhou$^{+}$}
\author{Wen Guan}
\author{Yiran Zhao}
\author{Chenjie Wang}
\author{Wenjian Wan}
\author{Sijia Yang}
\author{Zhenzhen Zhang}
\author{Chang Wang}
\author{J. C. Cao*}
\author{Heping Zeng*}
\author{Hua Li*}

\dedication{$^{+}$These authors contributed equally.}

\dedication{$^{*}$Corresponding authors. Email: hua.li@mail.sim.ac.cn; jccao@mail.sim.ac.cn; hpzeng@phy.ecnu.edu.cn.}

\begin{affiliations}
X. Liao, Dr. Z. Li, K. Zhou, W. Guan,  Y. Zhao, C. Wang, Dr. W. Wan, S. Yang, Dr. Z. Zhang, Prof. C. Wang, Prof. J. C. Cao, Prof. H. Li\\
Key Laboratory of Terahertz Solid State Technology, Shanghai Institute of Microsystem and Information Technology, Chinese Academy of Sciences, Shanghai 200050, China\\

X. Liao, K. Zhou, Y. Zhao, C. Wang, S. Yang, Prof. C. Wang, Prof. J. Cao, Prof. H. Li\\
Center of Materials Science and Optoelectronics Engineering, University of Chinese \\
Academy of Sciences, Beijing 100049, China\\

W. Guan\\
School of Information Science and Technology, ShanghaiTech University, 393 Middle \\
Huaxia Road, Shanghai 201210, China.\\

Prof. H. Zeng\\
State Key Laboratory of Precision Spectroscopy, East China Normal University, Shanghai 200062, China.

Prof. H. Zeng\\
Chongqing Key Laboratory of Precision Optics, Chongqing Institute of East China Normal University, Chongqing 401120, China.
\\

\end{affiliations}


\begin{abstract} 

Broadband dual-comb spectroscopy has attracted increasing interests due to its unique advantages in high spectral resolution, fast detection, and so on. Although the dual-comb technique is relatively mature in the infrared wavelengths, it is, currently, not commercially capable of practical applications in the terahertz regime due to the lack of high performance broadband terahertz dual-comb sources. In the terahertz frequency range, the electrically pumped quantum cascade laser (QCL) is a suitable candidate for the dual-comb operation. However, free running terahertz QCL dual-comb sources normally show limited optical bandwidths ($\sim$100-200 GHz). Although the resonant microwave injection locking has been widely used to broaden the emission spectra of terahertz QCLs by modulating the laser drive current at the cavity round-trip frequency, it is hard to be employed to broaden the dual-comb bandwidths due to the large phase noise induced by the resonant injection and non-ideal microwave circuits. Therefore, it is challenging to obtain broadband terahertz dual-comb sources that can fully exploits the laser gain bandwidth. Here, we employ an off-resonant microwave injection to significantly broaden the dual-comb bandwidth of a terahertz QCL dual-comb source emitting around 4.2 THz. The measured optical dual-comb bandwidth is broadened from 147 GHz in free running to $>$450 GHz under the off-resonant injection. The broadened dual-comb bandwidth is experimentally proved by the transmission measurements of a filter and a GaAs etalon. By performing a simple numerical analysis based on a rate equation model, we explain that the broadband dual-comb operation under the off-resonant microwave injection could be resulted from a wider lasing bandwidth and a higher degree of phase matching.

\end{abstract}

~\\


\section{Introduction}

An optical frequency comb is a coherent radiation source consisting of a series of equally spaced frequency lines \cite{DelNature,Yasui2006APL,Udem2002Nature,Faist2016FC,Picque2019NP}. Due to its high stabilities in the frequency and time domains, the frequency comb has been widely used in various applications, e.g., spectroscopy, imaging, communications, and so on \cite{Cappelli2019NP,Keilmann2004OL,Yang2017OL,NearField,boonruangkan2021coherence}. Among them, the dual-comb technique is a direct application of the frequency comb, and shows prominent advantages in the fast spectroscopy without a need of any moving part. In comparison, traditional infrared and terahertz spectrometers (for instance, the Fourier transform infrared and the time domain spectrometers) need mechanical scanning mirrors for obtaining a complete spectrum \cite{Coddington2016DCS,VillaresNC,Sterczewski2020ACSPho}.

Extending the spectral coverage of frequency comb and dual-comb sources from infrared or visible wavelengths to other frequency ranges has been a key for fundamental research and industrial applications. The terahertz wave attracts progressive interests in last decades due to its unique characteristics, e.g., finger prints of vibrational and rotational absorptions of molecules, excellent transparency to many package materials, large bandwidth for communications, etc. However, the development of frequency comb and dual-comb sources in the terahertz frequency range is, currently, immature due to the lack of efficient terahertz radiation sources. The electrically pumped, semiconductor-based terahertz quantum cascade laser (QCL), with high output power, good far-field beam quality, and wide frequency coverage, is an ideal candidate for the frequency comb and dual-comb operation \cite{Rosch2015Octave,Barbieri2011NP,FirstTHzQCL,HighPower2}. Regarding the terahertz QCL dual-comb operation, different architectures have been experimentally demonstrated, for instance, on-chip dual-comb, dual-comb spectroscopy using a fast detector, compact dual-comb spectrometer employing a self-detection scheme \cite{Rosch2016Onchip,Scalari2019,Yang2016Optica,LiACSPhoton,Sterczewski2019Optica,Li2019Onchip}. However, the measured optical bandwidth for terahertz QCL dual-comb sources, so far, is rather limited (less than 200 GHz). Although a microwave double injection has been used to broaden the optical bandwidth of a terahertz QCL on-chip dual-comb system, the broadening effect is not significant because the strong resonant injection introduces large phase noise and distorts the dual-comb spectra. Therefore, it is still challenging to obtain a broadband terahertz dual-comb source using the current radio frequency (RF) techniques. 

Here, in this work we demonstrate an off-resonant microwave injection to significantly broaden the terahertz dual-comb bandwidth without bringing the sizable phase noise. The two terahertz QCL combs are fully separated. One of the lasers is used as a fast detector for the dual-comb detection. We show that the off-resonant microwave injection can significantly broaden the dual-comb bandwidth. The increased dual-comb bandwidth is experimentally proved by further measurements, i.e., transmissions of a known absorber and a GaAs sample. The stability of the dual-comb source is evaluated by the phase noise measurement, the frequency and amplitude Allen deviation analysis. Finally, a numerical simulation based on a rate equation model and a group velocity dispersion analysis is implemented to reveal the underlying physical reason for the broadened dual-comb bandwidth induced by the off-resonant injection.

\section{Experimental setup and free running dual-comb operation}

\begin{figure}[!b]
	\centering
	\includegraphics[width=0.9\linewidth]{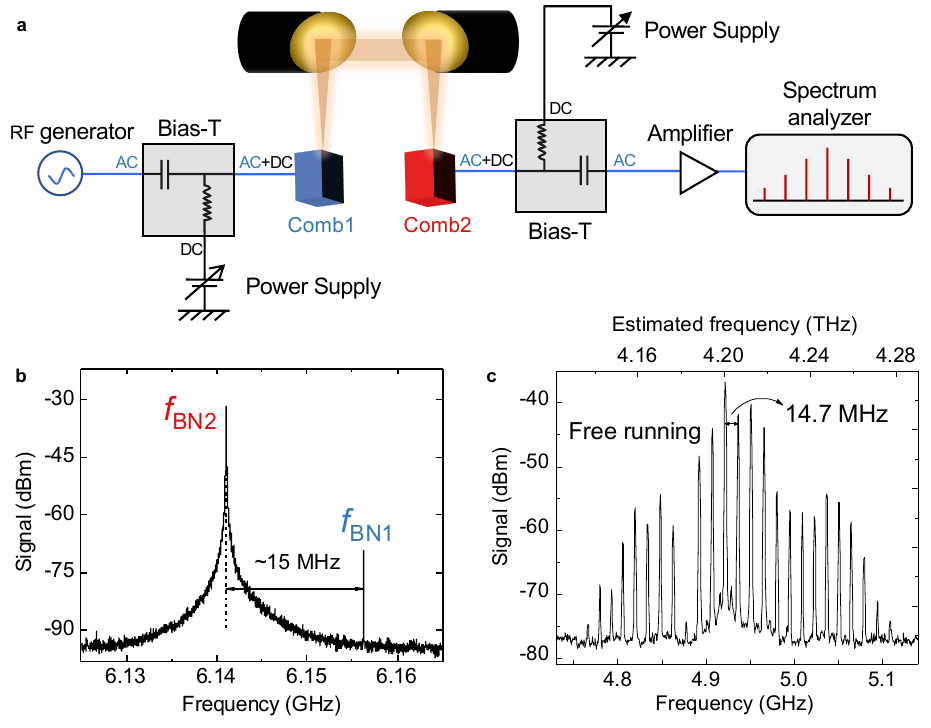}
	\caption{Experimental setup of the terahertz dual-comb source and dual-comb operation in free running. a) Experimental setup of the terahertz QCL dual-comb source with the microwave injection. Comb1 and Comb2 are terahertz QCLs with a ridge width of 150 $\mu$m and a cavity length of 6 mm. The RF generator injects microwave signals onto Comb1 via a Bias-T. The inter-mode beatnote and dual-comb signals are measured using Comb2 as a fast terahertz detector. b) Free running inter-mode beatnote signals measured using Comb2 as a detector with a resolution bandwidth (RBW) of 20 kHz and a video bandwidth (VBW) of 200 Hz. $f$$_{\rm{BN1}}$ and $f$$_{\rm{BN2}}$ are the inter-mode beatnotes of Comb1 and Comb2, respectively. The difference between $f$$_{\rm{BN1}}$ and $f$$_{\rm{BN2}}$ is around 15 MHz. c) Dual-comb spectrum measured when the two lasers are operated in free running. The measured line spacing is 14.7 MHz. The dual-comb spectrum was recorded with a RBW of 1 MHz and a VBW of 5 kHz. The results shown in (b) and (c) are recorded as Comb1 and Comb2 are operated at 990 mA at 30.1 K and 985 mA at 28.4 K, respectively.}
	\label{Schematic}
\end{figure}

 Figure \ref{Schematic}a shows the experimental setup of the off-resonant injection terahertz QCL dual-comb source (see Figure S1 in Supporting Information for a photograph of the experimental apparatus). The two QCLs are based on a hybrid active region design and processed into a single plasmon waveguide geometry (see Experimental Section). The dual-comb operation requires a slight difference between the two inter-mode beatnotes, $f_{\rm{BN1}}$ and $f_{\rm{BN2}}$. However, $f_{\rm{BN1}}$ and $f_{\rm{BN2}}$ cannot differ too much. Otherwise, the generated dual-comb signal will be out of the detection bandwidth. In this work, to obtain an acceptable difference between $f_{\rm{BN1}}$ and $f_{\rm{BN2}}$ (tens of MHz), a nominally identical laser cavity length of 6 mm is cleaved for both lasers. Due to the imperfections in cleaving and device fabrication processes, the two lasers, in principle, are not exactly identical, which finally results in a slight difference between $f_{\rm{BN1}}$ and $f_{\rm{BN2}}$. The laser ridge width of 150 $\mu$m is selected because it is more favorable for the frequency comb operation \cite{Zhou2019APL}. As shown in Figure \ref{Schematic}a, the microwave injection is directly applied onto Comb1. The injection conditions, i.e., resonant or off-resonant injections, and RF power, can be easily tuned by controlling the RF generator. In order to change the terahertz light coupling between Comb1 and Comb2, two parabolic mirrors are used. The beating of the two laser combs is self-detected using Comb2 as a detector. Note that the two laser combs have same functions, and therefore, the other laser (Comb1) can be also used as a detector for the dual-comb detection. Because the electron relaxation time in terahertz QCLs is fast in picoseconds, the QCL itself can be used as a fast detector for the multiheterodyne dual-comb detection with a bandwidth up to tens of gigahertz \cite{Li2015OE}. The RF signals including the inter-mode beatnote and dual-comb spectra are sent to the AC port of  the Bias-T that is connected to Comb2, and then amplified using a microwave amplifier with a gain of 30 dB, and finally registered on a spectrum analyzer.

To prove the dual-comb operation, we show the inter-mode beatnotes of Comb1 and Comb2 ($f$$_{\rm{BN1}}$ and $f$$_{\rm{BN2}}$), and the dual-comb spectrum obtained without any microwave perturbations in Figures \ref{Schematic}b and \ref{Schematic}c, respectively. In free running, when the two lasers are electrically pumped at currents of 990 mA (Comb1) and 985 mA (Comb2), two narrow inter-mode beatnote lines ($f$$_{\rm{BN1}}$ and $f$$_{\rm{BN2}}$) can be obtained with a frequency difference of $\sim$15 MHz. This frequency difference determines the line spacing of the dual-comb spectra. Indeed, as shown in Figure 1c, the measured dual-comb spectrum centered around 4.95 GHz demonstrates a line spacing of 14.7 MHz which is in agreement with the measured frequency difference as shown in Figure \ref{Schematic}b. Note that the 300 kHz mismatch is due to the frequency drift during the measurement. The dual-comb spectrum in Figure \ref{Schematic}c spans over 357 MHz with 25 lines, which corresponds to an optical bandwidth of $\sim$140 GHz by considering the repetition frequency of the laser combs. The dual-comb setup shown in Figure \ref{Schematic}a is flexible for the optical coupling tuning compared to the experimental setup used in ref. \cite{LiACSPhoton} where two lasers are mounted on a same sample holder and the optical coupling between the two lasers cannot change once they are mounted. As shown in Figure S2 in Supporting Information, by changing the position of Comb2 along the optical axis, we are able to tune the optical coupling between the two laser combs. It can be clearly seen that the inter-mode beatnotes ($f$$_{\rm{BN1}}$ and $f$$_{\rm{BN2}}$), dual-comb line number, dual-comb line spacing can be tuned by changing the optical coupling.

\section{Results and discussion}

\subsection{Dual-comb under off-resonant microwave injection}

\begin{figure*}[!t]
	\centering
	\includegraphics[width=0.95\linewidth]{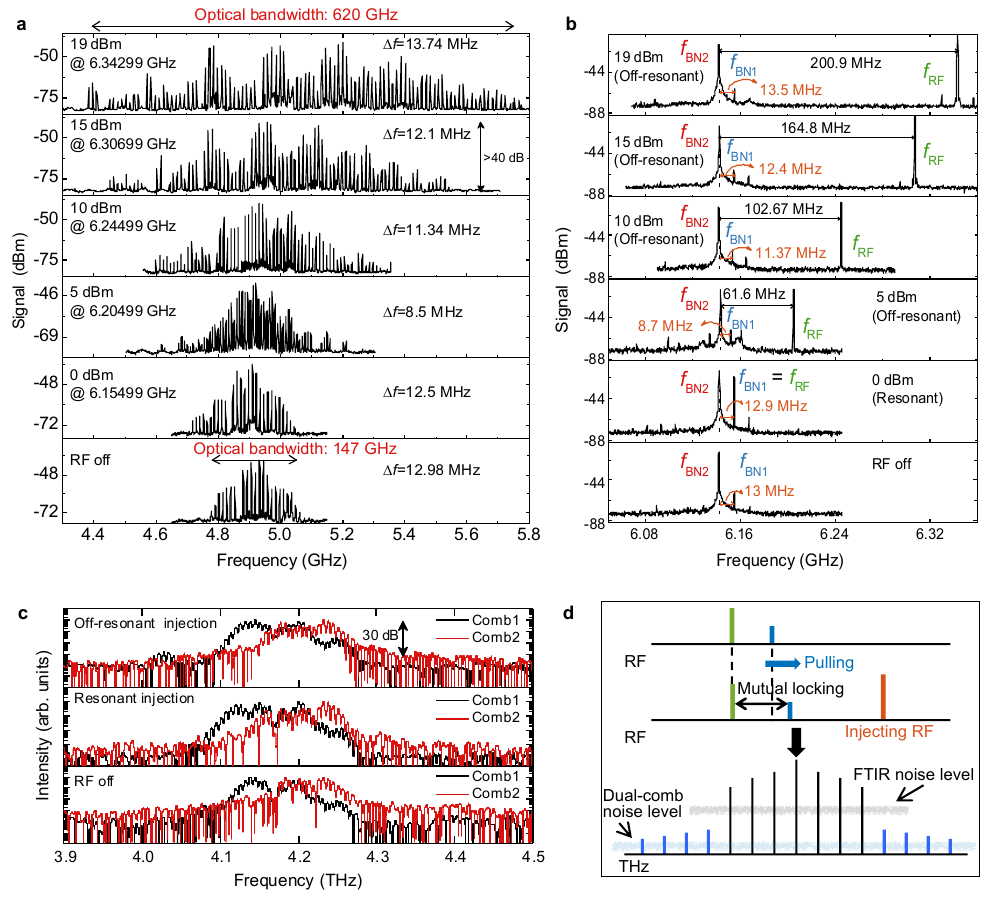}
	\caption{Broadband dual-comb operation under the off-resonant microwave injection. a) Dual-comb spectra measured with various microwave injection conditions. As a reference, the dual-comb spectrum in free running (RF off) is shown in the lowest panel. From bottom to top, the injection condition changes from low power resonant injection to high power off-resonant injection. All data were recorded with a RBW of 1 MHz and a VBW of 5 kHz. $\Delta$$f$ represents the measured dual-comb line spacings. b) Corresponding inter-mode beatnote and injecting RF signals ($f$$_{\rm{RF}}$) measured with a RBW of 300 kHz and a VBW of 20 kHz. c) Emission spectra of the two terahertz QCLs measured in different injection conditions. Resonant injection: 0 dBm @ 6.15615 GHz for Comb1; 0 dBm @ 6.14212 GHz for Comb2. Off-resonant injection: 19 dBm @ 6.34299 GHz for both laser combs. The spectral resolution of 0.08 cm$^{-1}$ is used for the emission measurement. d) Upper panel: Schematic demonstration of the two injection mechanisms, i.e., injection locking (pulling) and mutual locking, induced by the off-resonant injection. The blue and green lines represent $f$$_{\rm{BN1}}$ and $f$$_{\rm{BN2}}$, respectively, and the red line is the injecting microwave signal ($f$$_{\rm{RF}}$). Lower panel: Schematic plot of the comparison of noise levels of the FTIR and dual-comb measurements.}
	\label{injection}
\end{figure*}

Figure \ref{injection} demonstrates the effect of microwave injection on the terahertz dual-comb laser source. The experimental results obtained as the lasers are operated in free running (RF off) are also shown for a clear comparison. In Figure \ref{injection}a, we show the dual-comb evolution as the injection condition is changed. The resonant microwave injection at 6.15499 GHz with 0 dBm RF power is first applied, which shows limited effects on the dual-comb bandwidth compared with that obtained in free running. As the RF power is increased to 5 dBm, actually the dual-comb signal disappears immediately or signals with a broad pedestal are observed (see Figure S13d, Supporting Information). It indicates that high power RF injection brings about large phase noise, which damages the coherence of the terahertz modes. Similar phenomena were also found in an on-chip dual-comb source under a microwave injection \cite{Li2019Onchip}. In order to bring back the dual-comb signal, we introduce the off-resonant microwave injection (i.e., the injection frequency is far away from the repetition frequencies of the comb lasers) when the RF power is greater than 5 dBm. It can be clearly seen from Figure \ref{injection}a that by implementing the off-resonant microwave injection, we obtain not only  stable dual-comb operation but also broad dual-comb bandwidths. Especially, when we inject at 6.34299 GHz ($\sim$200 MHz away from $f_{\rm{BN2}}$) with a 19 dBm power, the optical bandwidth can reach 620 GHz with more than 90 lines, which is more than 3 times larger than the optical bandwidth measured in free running (147 GHz). To prove that the measured broad dual-comb bandwidth is resulted from the optical coupling of the two laser combs rather than some fake signals from electronic crosstalks, in Video 1 (Supporting Information) we show the real-time dual-comb operation when a metal board is used to block the terahertz beams. It can be clearly seen that when the optical coupling is blocked, the dual-comb signal disappears immediately.

Different from the resonant injection where the inter-mode beatnote frequency (or repetition frequency) is firmly locked at the injecting microwave frequency and the locking range is in a level of dozens of megahertz depending on the power of the injecting microwave signal (see Figure S3 in Supporting Information), the off-resonant injection cannot firmly lock the laser beatnote to the injecting frequency because the two frequencies are far away from each other. As shown in Figure \ref{injection}b, under the off-resonant injection condition, $f_{\rm{BN1}}$ and  $f_{\rm{BN2}}$ are still around 6.15 GHz and the difference between them determines the dual-comb line spacing (see Figure \ref{injection}a). Although the off-resonant injection cannot firmly lock the beatnote frequency, it still affects the beatnotes $f_{\rm{BN1}}$ and  $f_{\rm{BN2}}$. In Figure S4 (Supporting Information), the evolution of $f_{\rm{BN1}}$ and  $f_{\rm{BN2}}$ with the change of injection condition is plotted. Under the off-resonant injection, with the increase of the RF power, an obvious pulling effect can be clearly observed, which is a sign that the locking mechanism starts to work. From Figure \ref{injection}b, we can also observe that the sidebands beside the two beatnotes becomes stronger as the RF injection is applied, which indicates an increased interaction (mutual locking) between the two beatnotes. Therefore, the off-resonant injection can, in principle, partially lock the two inter-mode beatnotes, $f$$_{\rm{BN1}}$ and  $f$$_{\rm{BN2}}$. The data shown in Figures \ref{injection}a and \ref{injection}b were recorded using Comb2 as a detector. As we already explained previously, in principle, the other laser can be also used as a detector for the dual-comb detection. In Figure S5 (Supporting Information), we show the dual-comb and corresponding beatnote spectra measured using Comb1 as a detector under different injection conditions. We further show that as the power of the resonant injection is increased to a level (14 dBm), the coherence of the comb operation will be damaged and then the dual-comb spectrum disappears immediately (see Video 2 in Supporting Information). Because each laser has slightly different detection capability and the optical coupling strength can be different when using different lasers as detectors, we see the dual-comb results shown in Figure \ref{injection}a and Figure S5a in Supporting Information are not identical.

In Figure \ref{injection}c, we show the terahertz emission spectra of the laser combs under different conditions, i.e., RF off (lower panel), resonant injection with 0 dBm power (middle panel), and off-resonant injection with 19 dBm power (upper panel), measured using a Fourier transform infrared (FTIR) spectrometer (see Experimental Section). Compared to the emission spectra measured in free running (RF off), a slight spectral broadening can be observed once the microwave injection (either resonant or off-resonant) is applied. The lasing region is measured between 4.1 and 4.3 THz with a lasing bandwidth of 200 GHz. The results shown in Figure \ref{injection}c indicate that the microwave injection can broaden the emission spectra of the QCL combs. However, the lasing bandwidth measured from the FTIR is far less than the optical bandwidth of 620 GHz measured from the dual-comb spectrum shown in Figure \ref{injection}a. The bandwidth discrepancy between the FTIR and the dual-comb measurements can be explained as follows. First of all, in the two measurements the detectors show different performances. In the FTIR setup, a far-infrared deuterated triglycine sulfate (DTGS) detector was used for the detection of the terahertz signal and the noise equivalent power (NEP) of the detector is in a level of nW/Hz$^{1/2}$. While, the QCL detector demonstrates a NEP in tens of pW/Hz$^{1/2}$ \cite{Li2019Onchip}, which shows much stronger abilities for detecting weak signals than the DTGS detector. Therefore, some generated terahertz modes resulted from the off-resonant injection that are below the noise level of the FTIR can be still detectable by using the dual-comb setup. On the other hand, it is worth noting that the QCL configurations in the two measurements are different. In the dual-comb measurement, the two terahertz QCLs are optically coupled (see Figure \ref{Schematic}a); while in the FTIR measurement, each time only one QCL is switched on and its light is directly sent to the FTIR chamber for spectral measurements. Although pump conditions, i.e., drive current and RF injection, are identical for the dual-comb and FTIR measurements, the optical feedback situations in the two measurements are entirely different. We are inclined to believe that the optical feedback in the dual-comb measurement geometry contributes to the broadening of the optical bandwidth. Because of the above-mentioned two reasons, we observe much wider optical bandwidth from the dual-comb than that from the FTIR.

Here, we give more elaborations on the broadening mechanisms of the dual-comb bandwidth under the off-resonant microwave injection. As shown in Figure \ref{injection}d, two locking mechanisms resulted from the off-resonant injection can, in principle, contribute to the significant bandwidth broadening. The first one is the direct injection locking of $f_{\rm{BN1}}$, which is known as a powerful tool to broaden the emission spectra of QCLs\cite{Li2015OE,Wang2009OE,SirtoriLPR}. This injection locking mechanism can be proved by the pulling effect observed by increasing the injecting RF power (see Figure \ref{injection}b and Figure S3 in Supporting Information). As the power of the injecting RF is increased, we can see $f_{\rm{BN1}}$ is pulled by the injecting RF signal, which is the phenomenal sign of the start of injection locking\cite{Li2019Onchip,Gellie2010OE}. The other mechanism is the mutual locking effect, which can be often observed in the optical injection locking experiment where two optical frequencies with a slight frequency difference inject with each other and finally the two frequencies will be locked together. Indeed, when the two lasers are completely locked together, the dual-comb signals will be destroyed. But in our case, due to the relatively low optical power injection, we prefer to say a partial locking, which refers to the mutual interaction and the two inter-mode beatnotes are still different. Here as shown in Figure \ref{injection}b, even in free running mode (RF off), we can see the generated weak sidebands on the right of $f_{\rm{BN1}}$ and on the left of $f_{\rm{BN2}}$, which is resulted from the optical interactions of the two terahertz beams. Furthermore, as the external RF signal is applied and the RF power is increased gradually, we can see the generated sideband on the right side of $f_{\rm{BN1}}$ becomes stronger. It indicates that the mutual locking mechanism takes effect as the RF injection is applied. Different from the traditional optical injection locking, the mutual locking doesn't lock the two repetition rates ($f_{\rm{BN1}}$ and $f_{\rm{BN2}}$) together and we can still observe two separate repetition frequencies as shown in Figure \ref{injection}b. The two locking mechanisms, i.e., injection locking and mutual locking, function simultaneously and finally result in the broadening of the emission spectra as shown by the blue lines in Figure \ref{injection}d. Due to the optical bandwidth discrepancies between FTIR and dual-comb measurements (see Figures \ref{injection}a and \ref{injection}c), we assume that the mutual locking that is absent in the FTIR measurement plays a vital role in the bandwidth broadening. It is worth noting that for either the injection locking or mutual locking, the deep-level mechanism for the broadening is the four-wave mixing locking. The four-wave mixing is the inherent and important mechanism to generate equidistant frequency lines in the nonlinear GaAs gain medium. We assume that both injection locking and mutual locking can largely enhance the four-wave mixing locking effect in the QCLs, which then significantly broaden the emission spectra of the QCL combs.

Note that one can assume that the broadening of the dual-comb spectra under the off-resonant microwave injection is resulted from the nonlinearity of the RF components rather than from the optical beating, due to the presence of the inter-mode beatnote sidebands observed in Figure \ref{injection}b. In order to prove that the RF nonlinearity is not responsible for the dual-comb broadening, we performed another dual-comb measurement using two different QCL combs. The two new QCLs have the same active region as the ones used in Figure \ref{injection} but the laser cavity length is 5.5 mm which is slightly shorter than the ones shown in Figure \ref{injection}. The measured dual-comb and inter-mode beatnotes in different situations are shown in Figures S6 and S7 (Supporting Information). It can be clearly seen that under the off-resonant injection, the dual-comb spectra are significantly broadened, while the corresponding inter-mode beatnote spectra don’t show sidebands (see top panels of Figures S6 and S7, Supporting Information). If the RF nonlinearity takes effect in the broadening of the dual-comb spectra, the sidebands of the inter-mode beatnotes are expected to be observed. Therefore, the results shown in Figures S6 and S7 (Supporting Information) indicate that the appearance of the inter-mode beatnote sidebands is not the reason for the broadening of the dual-comb spectra.

In addition to the RF sidebands, we also investigate the possibility of terahertz sideband generation under the strong off-resonant injection. In Figures S8 and S9 in Supporting Information, we show the detailed multiheterodyne beating processes when the terahertz sidebands are involved. Three contradicts between the experimental result and the re-constructed spectrum, i.e., mode aliasing, amplitude variation, and dual-comb bandwidth, are observed. Therefore, we conclude that the generated terahertz sidebands under the strong off-resonant injection is not the reason for the broadening of the dual-comb bandwidth. The details for the comparison between the measured and re-constructed dual-comb spectra can be found in Figure S9 in Supporting Information.

\subsection{Experimental proofs of the broad dual-comb bandwidth}

To further prove that the broad dual-comb bandwidth under the off-resonant microwave injection is resulted from the mode proliferation in the terahertz range, two experiments are carried out by employing the dual-comb setup, i.e., transmission measurements of a known terahertz filter and a GaAs etalon.

\begin{figure*}[!t]
	\centering
	\includegraphics[width=0.9\linewidth]{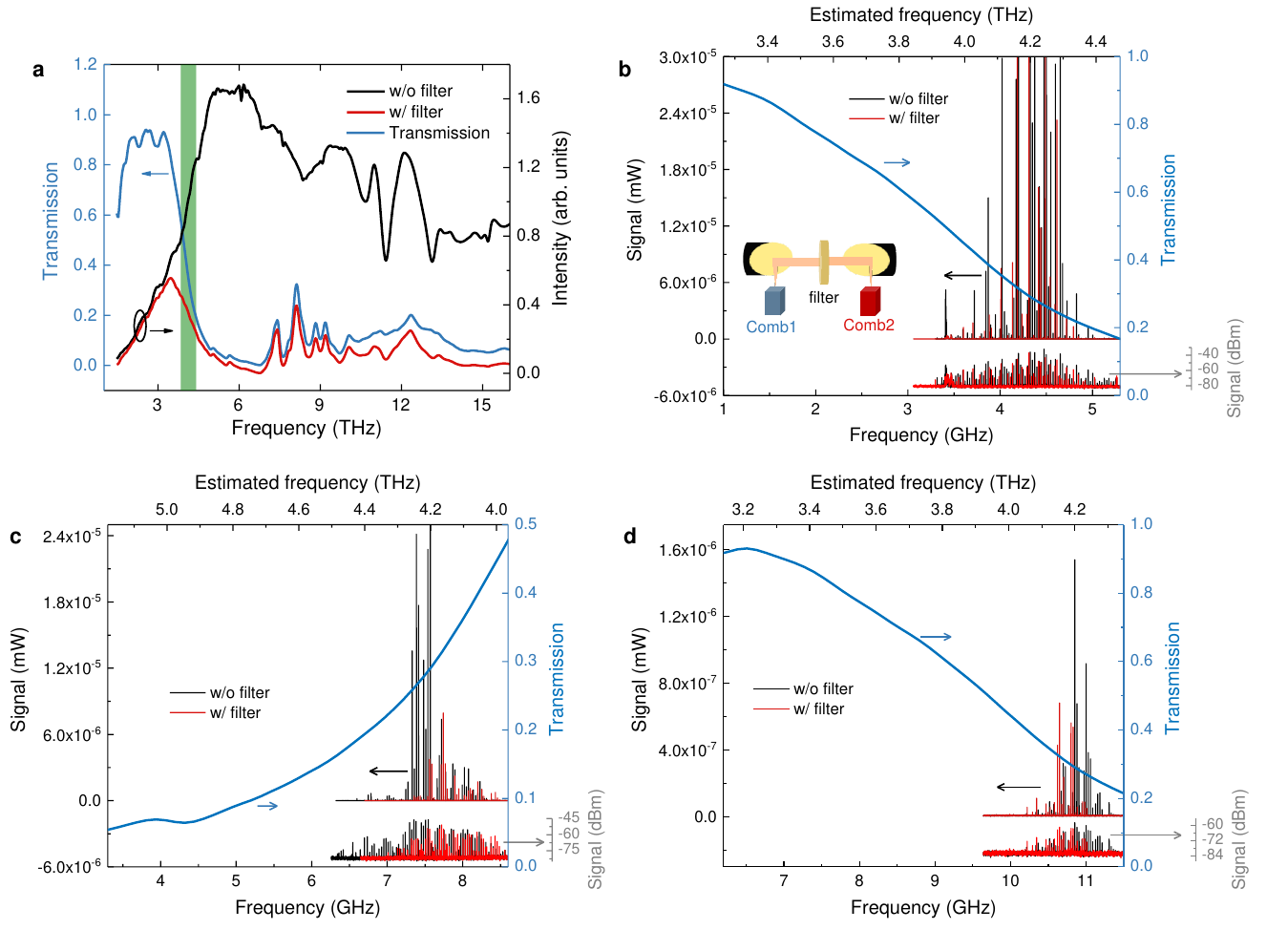}
	\caption{Dual-comb measurements of a filter. a) Transmission spectrum (blue curve) of the selected bandpass filter obtained by taking ratios of the data of the red curve to those of the black curve. All data were recorded using a FTIR spectrometer. The green shaded area approximately indicates the operation frequency range of the QCL combs (3.8-4.4 THz). b), c), d) are dual-comb spectra measured without (black curves) and with (red curves) the filter at different carrier frequencies in the microwave regime. Their carrier frequencies are around 4, 8, and 11 GHz in b), c), and d), respectively. The left y-axes show the dual-comb signal in a unit of mW, while the grey right y-axes show the dual-comb power in a unit of dBm. The top x-axes show the estimated frequencies in THz which correspond to the frequencies in microwave (bottom x-axes). The blue curves which are bounded to the blue right y-axes and top x-axes show the measured transmission spectra of the filter for comparison. For measurements shown in b)-d), the data were recorded with a RBW of 100 kHz as Comb1 is operated at 1070 mA at 26 K and Comb2 is operated at 1150 mA at 30 K. Comb1 is under the off-resonant injection situation (see the top panel of Figure \ref{injection}a). The inset of b) is the schematic of the experimental setup for the filter transmission measurement.}
	\label{filter}
\end{figure*}

Figure \ref{filter} shows the dual-comb measurements of a filter. The blue curve in Figure \ref{filter}a is the measured transmission (using FTIR) of the filter used in this study by taking ratios of the data of the red curve to those of the black curve. It can be seen that the filter is not narrow-band and its bandpass region is not precisely located in the optical range of the QCL combs. However, the high frequency tail of the filter fully covers the operation frequency range of the QCLs from 3.8 to 4.4 THz (see the shaded area in Figure \ref{filter}a). The slope showing the decrease in transmission with increasing frequency can be employed to identify if the generated comb lines are real or fake. The main idea is that when the filter is placed in the beam path of the dual-comb setup, the dual-comb spectra will be affected due to the frequency-dependent absorption (or transmission) of the filter. By comparing the dual-comb spectra without and with the filter, we, in principle, can prove if the dual-comb lines are generated from the beats of the modes in the terahertz range.
		
To perform the measurement, Comb1 is under the off-resonant injection situation. We first record the dual-comb spectra using Comb2 as a detector with the beam path purged with nitrogen and then record the dual-comb spectra with the filter placed in the beam path (see the inset of Figure \ref{filter}b) for comparison. The main results are shown in Figures \ref{filter}b, \ref{filter}c, and \ref{filter}d for the dual-comb spectra recorded at carrier frequencies around 4, 8, and 11 GHz, respectively. Because different beats of the two QCL comb lines can generate dual-comb signals at different carrier frequencies, in Figures \ref{filter}b-\ref{filter}d, we show three sets of dual-comb spectra at three lowest microwave carrier frequencies. As we already illustrated in ref. \cite{LiACSPhoton}, the microwave-terahertz frequency link for two neighbouring sets of dual-comb lines shows opposite trend, i.e., if for one set of dual-comb lines, the increase in microwave frequency corresponds to an increase in terahertz, then for the neighbouring sets of dual-comb lines, the increase in microwave frequency will correspond to a decrease in terahertz. Because of this, one can find that the microwave-terahertz link shown in Figure \ref{filter}c shows opposite trend as those shown in Figures \ref{filter}b and \ref{filter}d. If the broadening of the dual-comb is due to the RF nonlinearity rather than the broadening of the terahertz modes, the “symmetric absorption” behaviour shown in Figures \ref{filter}b-\ref{filter}d will not be observed. To help analyze the data, we plot the transmission of the filter as a reference in Figures \ref{filter}b-\ref{filter}d. It can be clearly seen that for all cases, the dual-comb results show good agreements with the measured transmission of the filter. In the higher terahertz frequency tail, the dual-comb lines are strongly attenuated or even removed due to the stronger absorption (or smaller transmission). Therefore, the dual-comb measurements of the filter can qualitatively prove that the significant broadening of the dual-comb bandwidth is resulted from the proliferation of the terahertz modes due to the off-resonant microwave injection. In Figure S10 (Supporting Information), we show the corresponding dual-comb measurement of the filter when the two terahertz QCLs are operated in free running mode.

\begin{figure*}[!b]
	\centering
	\includegraphics[width=0.95\linewidth]{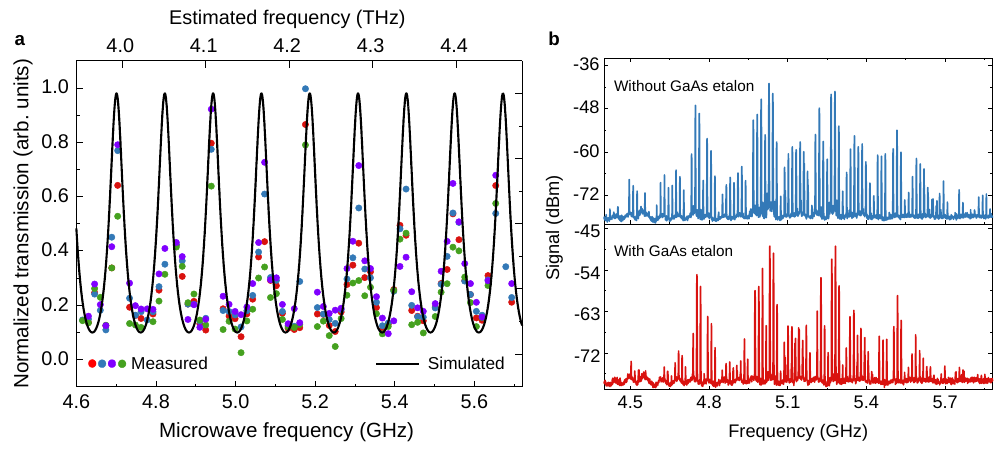}
	\caption{Transmission of a GaAs etalon sample measured using the terahertz dual-comb source under the off-resonant injection. a) Transmission spectrum of a GaAs etalon sample (625 $\mu$m $\pm$ 25 $\mu$m). The circles (bottom axis) with different colors are experimental results measured at different times while the black curve (top axis) is a fit. b) Dual-comb spectra with and without the GaAs etalon sample. The off-resonant injection at 6.34299 GHz with a power of 19 dBm is applied onto Comb1.}
	\label{etalon}
\end{figure*}

Figure \ref{etalon} shows the transmission of a GaAs etalon sample measured by employing the dual-comb source under off-resonant injection. To perform the measurement, we just simply insert the sample in the parallel beam between the two parabolic mirrors (replacing the filter shown in the inset of Figure \ref{filter}b) and then record the dual-comb spectra in real-time. Figure \ref{etalon}a plots the normalized transmission of a 625 ($\pm$25) $\mu$m thick GaAs etalon. The dots are experimental results that are obtained by taking ratios of the peak intensities of the dual-comb spectra measured with and without the GaAs etalon sample (see Figure \ref{etalon}b). The black curve is the transmission fit calculated by considering a refractive index of 3.6 and a sample thickness of 720 $\mu$m. Note that the sample thickness obtained from the fitting is slightly thicker than the nominal value, which can be attributed to the sample tilt during the measurement.

From Figures \ref{filter} and \ref{etalon}, we can experimentally prove that the observed broadening of the dual-comb bandwidth under the off-resonant microwave injection is indeed resulted from the proliferation of the terahertz modes. However, we have to note that the transmissions of the filter or GaAs etalon obtained from the dual-comb measurements are not precisely coincident with the FTIR measurement or fitting. This is because that when we insert the elements (filter or etalon samples) in the parallel path between the two laser combs, the optical coupling between the lasers will be accordingly changed, which strongly impacts the measured dual-comb spectra. Furthermore, the change is not linearly proportional to the absorption of the sample at each frequency. Sometimes it also introduces the energy transfer between the different modes and shift of peak positions which can be clearly seen in Figures \ref{filter} and \ref{etalon}. Therefore, it is hard to accurately perform quantitative transmission measurements for solid samples employing the current dual-comb setup. However, the results shown in Figures \ref{filter} and \ref{etalon} are sufficient to prove the optical bandwidth of the comb lasers under the off-resonant microwave injection.

\subsection{Stability evaluation of the dual-comb signal}

\begin{figure}[!t]
	\centering
	\includegraphics[width=0.95\linewidth]{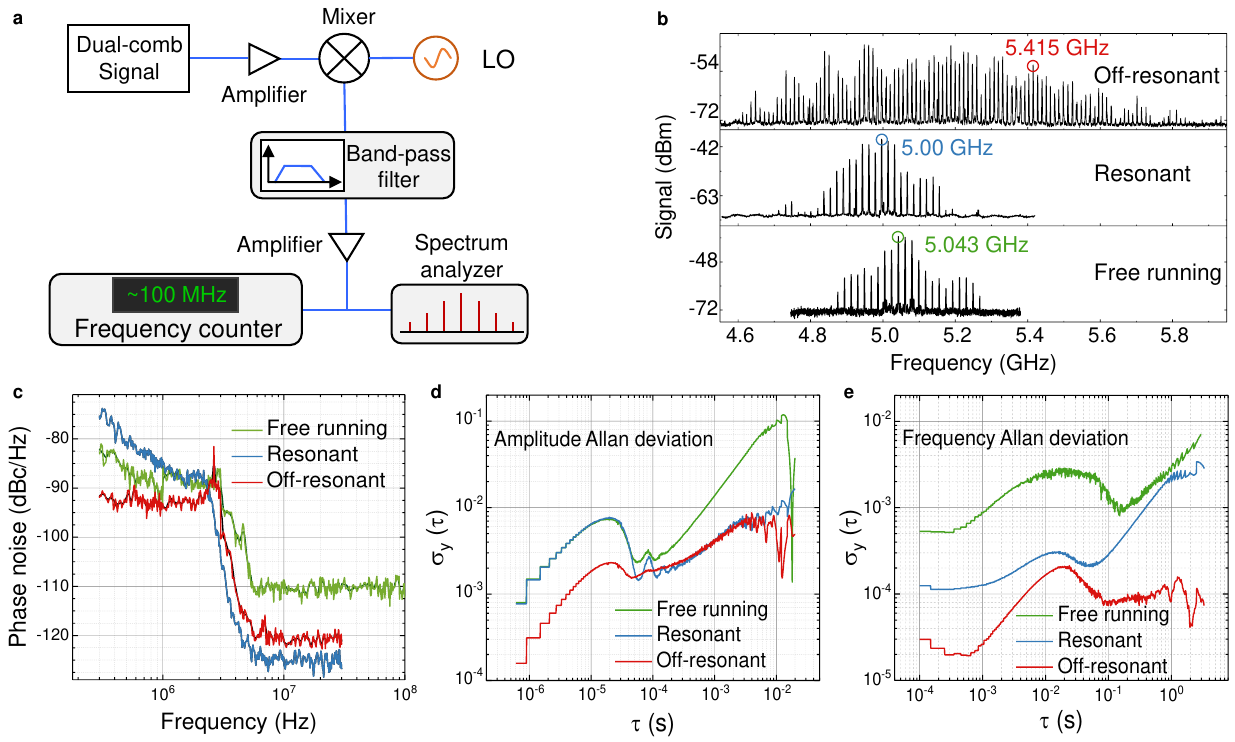}
	\caption{Stability evaluation of the dual-comb signals at different conditions. a) Experimental setup of the Frequency Allan deviation analysis for the dual-comb signals. To record the frequency as a function of time, the dual-comb signals should be converted to low frequencies within the bandwidth of the frequency counter. LO: Local oscillator signal (100 MHz difference from the selected signal). b) Dual-comb spectra measured with a RBW of 1 MHz in free running, resonant injection, and off-resonant injection. The circles mark the dual-comb lines that are selected for the stability evaluation. (c), (d) and (e) are measured phase noise spectra (RBW: 10\%, tolerance: 10\%), frequency Allen deviation, and amplitude Allen deviation of the lines marked in b, respectively. All the data were recorded when Comb1 and Comb2 are operated at 990 and 985 mA, respectively.}
	\label{stability}
\end{figure}

The stability of the dual-comb signals is important for practical applications. In Figure \ref{stability}, we evaluate the stability of the dual-comb signals obtained in different situations, i.e., free running mode, resonant injection and off-resonant injection. Figure \ref{stability}a plots the experimental setup for the frequency stability measurements. To meet the bandwidth requirement of the frequency counter, the dual-comb signal is down-converted from 5 GHz to 100 MHz and then only one dual-comb line is finally selected using a band-pass filter for the frequency stability measurement (see Experimental Section for details). In Figure \ref{stability}b, from bottom to top panels we show three dual-comb spectra recorded in free running, resonant injection and off-resonant injection situations. As indicated by the circles, three typical lines located at 5.043 GHz (free running), 5.00 GHz (resonant injection), and 5.415 GHz (off-resonant injection) are chosen for the stability evaluation. Figures \ref{stability}c, \ref{stability}d, and \ref{stability}e plot the phase noise spectra, amplitude Allan deviation, and frequency Allan deviation, respectively, for the three lines indicated by the circles in Figure \ref{stability}b. As expected, the phase noise and Allan deviation plots obtained in the injection condition show improved values than those measured in free running mode.

An effective way to further improve the stability of the dual-comb signal is to implement the phase locking. The complete locking of a dual-comb source needs all four frequencies (two repetition rates and two carrier offset frequencies) to be locked simultaneously, which will significantly complicates the system. Alternatively, as reported in ref. \cite{zhao2021active}, we can perform the locking of terahertz QCL dual-comb source by phase locking one of the dual-comb lines to a stable microwave oscillator without any control of the repetition rate and carrier offset frequency of individual laser comb. It has been shown that although just one dual-comb line is phase locked, the stability of other dual-comb lines is significantly improved. In view of this, here we also applied the locking technique onto the dual-comb source investigated here. The experimental results show that we can easily lock the repetition rate of one of the lasers (see Figure S11b and Video 3 in Supporting Information). However, the long-term locking of the dual-comb signal cannot be obtained. From Video 4 in Supporting Information, we see that the phase locking of the dual-comb source can work for a short period (around 9 seconds) then the locking fails for longer time. In ref. \cite{zhao2021active}, the two laser combs sit on a same sample holder in a cryostat, while in this work the two laser combs are completely separated and they don't share the same temperature fluctuations. Therefore, it is more difficult to obtain a stable phase-locking of the dual-comb source in this work. As can be seen from Video 4 in Supporting Information, the frequency drift of the dual-comb lines is almost on the borderline of the locking bandwidth of the phase locked loop (PLL). It, in principle, should be feasible to achieve the phase locking by further optimizing the Servo electronics to slightly increase the locking bandwidth of the PLL.

\subsection{Simulation}

\begin{figure*}[!t]
	\centering
	\includegraphics[width=0.9\linewidth]{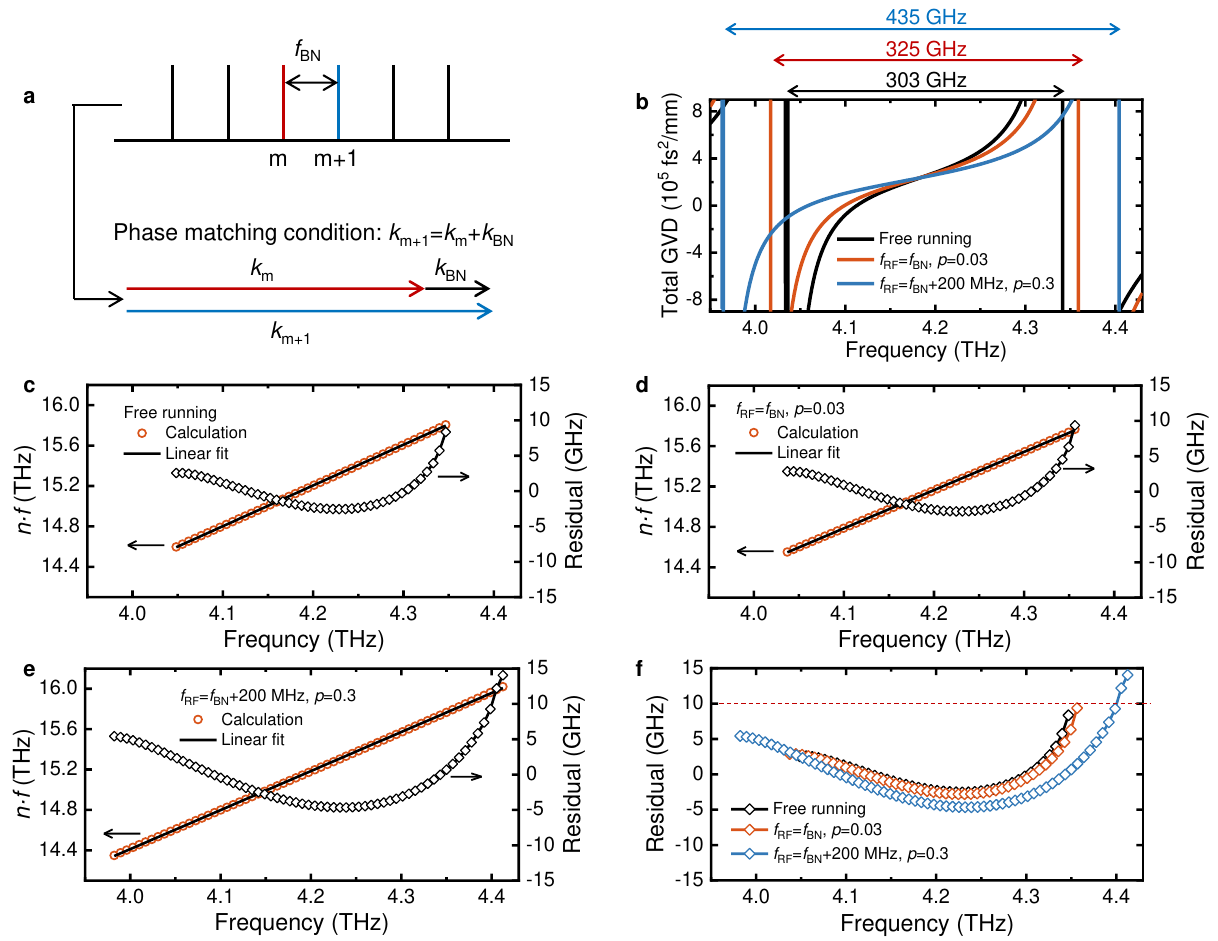}
	\caption{a) Phase matching condition for the m-th mode and (m+1)-th mode in the QCL frequency comb. b) Calculated total GVDs for the QCL operated in free running mode (black), resonant injection ($f_{\rm{RF}}$=$f_{\rm{BN}}$) with a low RF power ($p$=0.03) (red), and off-resonant injection ($f_{\rm{RF}}$=$f_{\rm{BN}}+$200 MHz) with a high power ($p$=0.3) (blue). As shown in b), the calculated lasing bandwidths of the QCL in free running, resonant injection, and off-resonant injection are 303, 325, and 435 GHz, respectively. c), d), and e) are calculated $n{\cdot}f$ as a function of $f$ for the laser operated in free running, resonant injection, and off-resonant injection, respectively. The black lines shown in c)-e) are the linear fits of the calculated results (red circles). The right y-axis shows the corresponding residuals in each panel. f) Comparison of the residuals for the QCL operated in free running (black), resonant injection (red), off-resonant injection (blue). The red dashed line indicates the level of residual of 10 GHz.}
	\label{phase matching}
\end{figure*}

We experimentally showed the broadening of the dual-comb bandwidth of the QCL dual-comb source under an off-resonant microwave injection. To understand the underlying reason of the broadening effect, we perform a simple simulation study. As it has been known that the frequency comb and dual-comb generation is resulted from the nonlinear processes, e.g., four-wave mixing, which are sensitive to the phase matching condition\cite{maker1962effects,wang2015generating}. And the phase mismatch should be reduced to minimum in order to obtain an efficient exchange of energy between the modes of a frequency comb\cite{jung2013optical}. In our simulation, we consider the m-th and (m+1)-th modes of a frequency comb. As shown in Figure \ref{phase matching}a, the phase matching condition is that the difference of the wave vectors of the two terahertz modes is equal to the wave vector of the beatnote frequency in the microwave range, which can be written as
\begin{equation}
k_{\rm{m+1}}=k_{\rm{m}}+k_{\rm{BN}}.
\label{pm1}
\end{equation}
By considering the refractive index, equation \ref{pm1} can be further written as
\begin{equation}
n(f_{\rm{m+1}}){\cdot}f_{\rm{m+1}}=n(f_{\rm{m}}){\cdot}f_{\rm{m}}+n(f_{\rm{BN}}){\cdot}f_{\rm{BN}},
\label{pm2}
\end{equation}
where $n$ and $f$ denote the refractive index and frequency, respectively, and $f_{\rm{BN}}$ is the inter-mode beatnote frequency which is equal to $f_{\rm{m+1}}$-$f_{\rm{m}}$. From equation \ref{pm2}, we can see that to satisfy the phase matching condition, the production of $n$$\cdot$$f$ should be linearly dependent on $f$ in the frequency comb region. Therefore, the calculation of $n$ is crucial for the phase matching. Note that the phase matching described in equations \ref{pm1} and \ref{pm2} is related to a second order nonlinearity. However, due to the comb nature, once the phase matching condition (equations \ref{pm1} and \ref{pm2}) is satisfied, the phase matching of the four-wave mixing, i.e., $k_{\rm{m}+1}=2 k_{\rm{m}}-k_{\rm{m}-1}$ ($k_{\rm{m}+1}$, $k_{\rm{m}}$, and $k_{\rm{m}-1}$ represent wave vectors of three consecutive lines of a comb), is naturally satisfied. Because of this, when we obtain the phase matching condition, the four-wave mixing locking is enhanced. Here in this study, we calculate $n$ as a function of $f$ from the total group velocity dispersion (GVD) of the QCL device. To calculate the GVD of the terahertz QCL under the off-resonant microwave injection, we start from a three-level rate equation analysis \cite{meng2012Optexpress} which is used to calculate the population inversion (${\Delta}N$) modulated by the injection signal. In the rate equation model, the alternating current (AC) density introduced by the RF modulation ($J_{\rm{RF}}$) directly superimposes on the DC density of the QCL device ($J_{\rm{DC}}$), which gives the total current density ($J$),
\begin{equation}
J=J_{\rm{DC}}+J_{\rm{RF}}
=J_{\rm{DC}}+p{\cdot}J_{\rm{DC}}{\cdot}{\sin}(2{\pi}f_{\rm{RF}}t),
\label{Jcurrent}
\end{equation}
where \textit{p} is a ratio of the induced current change by RF injection to the device current in free running mode\cite{wang2015active}. Therefore, the parameter $p$ is proportional to the RF injection power.

After the necessary parameters, i.e., $J$, laser dimensions, losses, relaxation times, etc., are introduced into the rate equation model, ${\Delta}N$ can be calculated. Note that for the calculation of ${\Delta}N$ at the off-resonant injection condition, a simple approximation is implemented. We assume that for a given RF power, the resonant injection brings about the largest ${\Delta}N$ and an attenuation in ${\Delta}N$ following a quadratic function is considered for the off-resonant injection. After we obtain ${\Delta}N$ which is a function of RF frequency ($f_{\rm{RF}}$) and RF power ($p$), we can then calculate the gain of the QCL using the Fermi's golden rule\cite{book1}. To obtain the lasing bandwidth, the gain clamping effect is taken into account. The threshold gain is calculated by $g_{\rm{th}}$=($\alpha_{\rm{w}}$+$\alpha_{\rm{m}}$)/$\Gamma$, where $\Gamma$=0.25 is the mode confinement factor, $\alpha_{\rm{w}}$ is the waveguide loss calculated from a two-dimensional finite-element modeling, and $\alpha_{\rm{m}}$ is the mirror loss. The calculated threshold gain  and (clamped) gain for the QCL operated in various conditions are shown in Figure S12 (Supporting Information). For all different cases shown in Figure S12, the threshold gain of the QCL doesn't changed but the gain profiles change, which results in the change of clamped gain bandwidth (or lasing bandwidth). Once the information of the clamped gain and waveguide loss is obtained, the total GVD is calculated by using the Kramers-Kronig relations, which includes the gain, waveguide, and material GVDs\cite{Zhou2019APL}. Note that in the simulation the gain GVD is calculated from the obtained clamped gain. Figure \ref{phase matching}b shows the calculated total GVDs for the QCL operated in different situations, i.e., free running, resonant injection ($f_{\rm{RF}}$=$f_{\rm{BN}}$) with a low RF power ($p$=0.03), and off-resonant injection ($f_{\rm{RF}}$=$f_{\rm{BN}}+$200 MHz) with a high power ($p$=0.3). It can be seen that under the off-resonant injection condition, the calculated total GVD shows a flatter frequency dependence, which indicates that the QCL with the off-resonant injection can be more suitable for frequency comb operation with wider frequency coverage. Once we obtain the total GVDs of the QCL under different operation conditions, the effective refractive index ($n$) can be calculated by solving the following differential equation
\begin{equation}
\textrm{GVD}=\frac{\partial}{\partial{\omega}}\,\frac{n(\omega)+\omega\frac{\partial{n(\omega)}}{\partial{\omega}}}{c},
\end{equation}
with $\omega$ and $c$ being the angular frequency and speed of light, respectively. Then, the product of $n{\cdot}f$ can be obtained to evaluate the phase matching condition. In principle, the linearity of the $n{\cdot}f$-$f$ plot directly reflects the degree of the phase matching. In Figures \ref{phase matching}c, \ref{phase matching}d, and \ref{phase matching}e, we show the calculated $n{\cdot}f$ as a function of $f$ for the QCL operated in free running, resonant injection, and off-resonant injection, respectively, in the corresponding lasing bandwidths. Here, a linear fitting is performed and the obtained residuals are used to evaluate the degree of the linearity or phase matching. Normally, the small residuals randomly distributed around 0 indicate the perfect phase matching. However, we can see in Figures \ref{phase matching}c-e that the three residual curves show a quadratic trend which indicates that the function $n{\cdot}f$ has some small nonlinearity\cite{hogg2005introduction}. The nonlinearity in frequency combs can be resulted from the dispersion\cite{gaeta2019photonic}. Figure \ref{phase matching}f plots the residuals of the QCL operated in the three different situations for a clear comparison. It can be seen that if a residual of 10 GHz is considered for the phase matching condition, the QCL under the off-resonant injection demonstrates a comb bandwidth of 410 GHz with the upper and lower frequency edges of 4.39 and 3.98 THz, respectively, while the calculated lasing bandwidth under the off-resonant injection is 435 GHz (see Figure \ref{phase matching}). The comparison shows that the phase matching condition shrinks the bandwidth by 25 GHz for the comb operation. A further evaluation of the comb bandwidth indicates that when compared to the free-running bandwidth, the off-resonant injection induced comb bandwidth enhancement in the lower frequency side is around 22\%, while in the higher frequency side the bandwidth is enhanced by 17\%. Moreover, it can be clearly seen that in the lower frequency side, the residuals obtained from the off-resonant injection are not much different from those calculated for the free-running QCL comb. But, in the higher frequency side, the residuals are obviously smaller than the free-running residuals. Therefore, we can say that both phase matching and lasing bandwidth broadening contribute to the comb bandwidth broadening under the off-resonant injection. And roughly speaking, 17\% and 22\% of the comb bandwidth broadening are resulted from the phase matching and lasing bandwidth broadening effect, respectively.

Here, we have to emphasize that the dual-comb system with microwave modulations is very complex, while the current simulation that is based on a rate equation model is relatively simple. Therefore, we, in principle, are not able to quantitatively reveal the modulation effects on the laser combs, especially for the bandwidth enhancement induced by the off-resonant injection. Although the calculated comb bandwidth enhancement is not exactly same with the measured one, the bandwidth broadening phenomenon resulted from the off-resonant injection can be revealed by the modeling. The quantitative evaluation of the dual-comb source can be improved by fully incorporating spatial hole burning and Kerr nonlinearity\cite{PhysRevLett2019Opacak} in the model. Furthermore, in the current model, although we considered the modulation effects of microwave frequency and power on the current of the laser (equation \ref{Jcurrent}), the interaction between the microwave frequency and the propagating beatnote frequency in the cavity was not taken into account. This approximation is reasonable when the injection power is low because we assume the injecting microwave signal cannot propagate along the laser cavity due to the strong attenuation in the single plasmon QCL waveguide (see also the texts in the next section). However, when a strong resonant microwave injection is applied ($f_{\rm{RF}}$=$f_{\rm{BN}}$ and $p$=0.3), the injected signal can partially interact with the inter-mode beatnote signal of the comb at the edge of the laser ridge. In Figure S13 (Supporting Information), we show the calculated results for the resonant injection with a high RF power ($p$=0.3) together with the results obtained under off-resonant injection with a high RF power ($p$=0.3). It can be seen that there is not much difference between the resonant and off-resonant injections with a high RF power in our model. However, in the experiment, we do observe significant difference between the two cases. The reason could be that in the experiment, when the QCL is resonantly modulated by a high RF power, the phase noise at the injection frequency which is proportional to the RF power is large. And the resonant interaction between the injecting RF signal and the generated inter-mode beatnote signal at the edge of the laser ridge is strong, which may result in large phase noise for the comb operation (see the broad pedestal in Figure S13d, Supporting Information) and phase mismatching. However, this interaction between the injecting RF signal and the generated beatnote signal at the edge of the laser ridge is not taken into account in our model. To the end, we have to clearly state that the current simple model cannot reveal the different phenomena between the off-resonant and resonant RF injections with high power.

\subsection{Discussion}

Concerning the phase matching between the terahertz waves and the microwave signal, one could question that in a single plasmon QCL, the microwave mode doesn't exist due to the strong signal attenuation. Our explanation is the following. Indeed, different from the metal-metal terahertz QCLs, the microwave signal in single plasmon lasers shows huge attenuation. However, it is worth noting that in our single plasmon QCLs, there are two microwave signals. One is the injecting microwave signal that is from the external RF generator. This signal is used to modulate the laser drive current and then the population inversion, which can alter the lasing bandwidth and group velocity dispersion (GVD) of the laser. The other microwave signal is the inter-mode beatnote signal generated by the beating of the terahertz modes. As reported in our recent paper\cite{Li2022real-time}, indeed the microwave transmission $S_{12}$ measurement shows that in a 15 mm long single plasmon terahertz QCL, the attenuation in the microwave frequency range from 0 to 40 GHz along the laser ridge is measured to be $\sim$2.6 dB/mm. This attenuation is attributed to the combined effect of the device conductance and free carrier absorption in the doped contact layers and metal contacts. Due to the strong absorption, the propagation of the microwave signal is heavily damped. However, in the QCLs the inter-mode beatnote signal is generated from the beating of the terahertz modes and the terahertz photons can propagate in the QCL cavity. Therefore, the inter-mode beatnote signal does exist in the entire cavity. We can say that the microwave signal and the terahertz wave co-propagate in the laser cavity. Because of this, the phase matching between the terahertz waves and the microwave can be obtained in the single plasmon terahertz QCLs.

It is worth mentioning that the bandwidth broadening induced by the off-resonant injection is reproducible. To prove its reproducibility, we performed another dual-comb experiment using a new pair of QCLs. The new lasers have the same active region structure as the ones used in the main paper and a different laser cavity length (5.5 mm). The results are shown in Figures S6 and S7 (Supporting Information). It can be clearly seen that under the off-resonant injection, we do observe the significant broadening of the dual-comb spectra. Because different lasers show different lasing and detection performances due to the imperfect growth and fabrication processes, the measured bandwidths for the new pair of lasers (Figures S6 and S7, Supporting Information) are not exactly identical to those shown in Figure \ref{injection}a. However, the significant broadening effect resulted from the off-resonant injection is observed and the reproducibility is verified.

Another general point that should be discussed here is how to select lasers to produce the broadband dual-comb signals under the off-resonant microwave injection. First of all, the experiments employing the single plasmon terahertz QCLs have been repeatedly performed many times and the broadening of the dual-comb bandwidth under the off-resonant microwave injection can be observed each time. Although we have not yet tested the metal-metal terahertz QCLs, we think the similar broadening effect can be observed if the phase matching condition can be optimized. The second important factor is the stability of the laser. As it is known, to perform the dual-comb measurement, a high level stability, characterized by narrow inter-mode beatnotes and low phase noise, should be obtained. Thirdly, the cavity length of the device is also a very important factor for the comb or dual-comb operation. We already did several experimental investigations in previous studies. In ref.\cite{Wan2017SR}, we compared the comb operation for terahertz QCLs with two different cavity lengths, i.e., 6 mm and 2.5 mm. We did observe that at most of drive currents, the short cavity laser showed broad inter-mode beatnote spectra (Figure S3 of ref.\cite{Wan2017SR}), which inferred that the short cavity laser cannot work as a frequency comb. If the comb operation cannot be obtained in the short cavity laser, the dual-comb bandwidth broadening won’t be observed under the RF modulation. The reason for the difference was that for the 6 mm cavity laser, the repetition frequency is much smaller, which means that dense modes are obtained. In this situation, the four-wave mixing locking can be stronger and therefore, much better comb operation is obtained. Furthermore, in our recent work\cite{Li2022real-time}, we experimentally show that a 15-mm long cavity terahertz QCL with a same active region is able to produce self-started optical pulses and a basic solitonic character of the long cavity QCL comb is observed. Although we have not yet finely tuned the cavity length to study the length effect on the frequency comb and dual-comb operation, the experimental results obtained from many devices with cavity lengths ranging from 2 to 15 mm show an empirical conclusion that the short cavity is not suitable for the QCL frequency comb operation for the given active region design. Finally, we have to emphasize that the comb measurements that translate the frequency from terahertz to microwave frequency range is high precision measurements, which can be affected by many factors. Due to the imperfections in the growth, fabrication and packaging processes, the environmental and electrical noises, and slight differences in the group velocity dispersion from chip to chip, even two devices with the same nominal cavity length will have different performances (refer to Figure 2a and Figure S5).

\section{Conclusion}

In summary, we have demonstrated a broadband terahertz QCL dual-comb operation under an off-resonant microwave injection. The optical bandwidth of the dual-comb source can reach $>$450 GHz under an off-resonant microwave injection with a power of 19 dBm. The increased dual-comb bandwidth induced by the off-resonant injection was experimentally proved by transmission measurements of a filter and a GaAs etalon. Finally, a simple numerical simulation based on a rate equation model was performed to partially explain the observed phenomena. The modeling showed that the enhanced lasing bandwidth and high degree of phase matching could be responsible for the significant dual-comb bandwidth broadening. However, it should be noted that the current simulation cannot fully reveal the underlying physical mechanisms of the observed results, especially for the case of the high power resonant microwave injection. In the future, for a deep and complete understanding of the work, more efforts should be followed. First of all, the model can be significantly improved by including more physical effects, such as, spatial hole burning and Kerr nonlinear effects, interaction between the RF signal of a synthesizer and the inter-mode beatnote signal of the comb, phase noise of the RF synthesizer especially at a high RF power condition, etc. On the other hand, further measurements can be carried out to directly verify the comb bandwidth under the off-resonant injection. For instance, the comb emission spectra with a dual-comb configuration can be measured with a much more sensitive terahertz detector.

\section{Experimental Section}

\threesubsection{Terahertz QCLs and emission spectral measurement}The terahertz QCL used in this work is based on a hybrid active region structure that combines the bound-to-continuum optical transitions and the resonant phonon scattering for depopulation of the lower laser state. The details of the layer structure can be found in ref. \cite{Wan2017SR}. The entire QCL active region is grown using a molecular beam epitaxy system on a semi-insulating GaAs (100) substrate. The grown wafer is then processed into single plasmon waveguide laser ridges. The ridges are cleaved into laser bars with cavity lengths ranging from 2 to 6 mm and then indium-soldered onto copper bases (heat sink). Finally, the QCL device is mounted onto a cold finger of a cryostat for electrical and optical measurements.

It is worth mentioning that in this work the deployment of the single plasmon waveguide can benefit the optical coupling between the two laser combs. It is known that the optical mode can go further into the substrate in the single plasmon waveguide QCL, while for the double-metal waveguide the mode is confined in the active region between two metal layers. Therefore, when the terahertz light emitted from one QCL is coupled into the other QCL (detection comb) via the device facet, the single plasmon waveguide can couple more light into the detector for the multiheterodyne dual-comb detection because it has much larger effective facet area for the optical interaction than the double-metal waveguide \cite{LiACSPhoton}. 

The emission measurement of the laser combs were performed as the lasers were driven in continuous wave (CW) mode at cryogenic temperature employing a Fourier transform infrared spectrometer (Bruker v80). To reduce the water absorption, the beam path in the spectrometer is evacuated down to 3 mbar and the beam path outside the spectrometer is purged with dry air. The spectra shown in Figure \ref{injection}c were recorded with a spectral resolution of 0.08 cm$^{-1}$ in a fast scan mode.

\threesubsection{Multiheterodyne dual-comb detection}The multiheterodyne dual-comb spectra shown in Figures \ref{Schematic}c, \ref{injection}a, and \ref{etalon}b were measured using the QCL comb itself as a fast detector. The experimental setup is shown in Figure \ref{Schematic}a. Both Comb1 and Comb2 emit terahertz light and simultaneously Comb2 can detect the multiheterodyne beating signals \cite{Rosch2016Onchip,LiACSPhoton,Li2019Onchip}. Since fast detectors working around 4 THz with a bandwidth up to the GHz level are not commercially available, the self-detection technique employed here can make the fast multiheterodyne detection feasible and largely simplify the experimental setup. To extract the RF signal, a Bias-T is connected to Comb2 (detector) as shown in Figure \ref{Schematic}a. The signal exits from the AC port of the Bias-T and then is amplified using a microwave amplifier, and finally the dual-comb spectra are registered on a spectrum analyzer (Rohde \& Schwarz, FSW26). Note that in the current geometry, the two lasers are equivalent which means that either laser can be used as the fast detector for the multiheterodyne detection. As shown in Figure S5 (Supporting Information), the dual-comb spectra were obtained using Comb1 as the detector. Although either laser can be used as the detector, the measured spectra using different detectors show different line numbers and signal to noise ratios. This is because that the optical coupling in reality cannot be the same when we use different lasers as detectors. Furthermore, the QCLs can demonstrate different detection abilities due to the non-ideal growth, fabrication processes, mounting procedures, etc.

\threesubsection{Microwave injection}The microwave injection is an efficient tool to lock the repetition rate of a semiconductor laser and then further broaden the emission spectrum, which has been widely used for mid-infrared and terahertz QCLs \cite{Li2015OE,Wang2009OE,Gellie2010OE}. In previous studies, the microwave signal is resonantly injected to a laser at its cavity round-trip frequency. In this work, we find that the resonant injection can only work at low RF power. As the RF power is increased to some level (for example, 5 dBm using Comb2 as detector and 14 dBm using Comb1 as detector), the resonant injection will result in large phase noise and the dual-comb signal disappears then (See Video 2 in Supporting Information). Therefore, here we employ the off-resonant injection technique when the RF power is large. To extract the high-frequency signal from the laser cavity, we use a microwave transmission line. The microwave injection is only applied onto one of the lasers, for example, Comb1 as shown in Figure \ref{Schematic}a. The microwave signal delivered from the RF generator (Rohde \& Schwarz, SMA100B) is sent to Comb1 via the AC port of a Bias-T. The signal frequency and power can be precisely tuned by the RF generator. By shifting the injecting frequency from $f$$_{\rm{BN1}}$, we can easily switch from resonant injection to off-resonant injection.

\threesubsection{Estimation of the frequency link between microwave and terahertz} In the current dual-comb system, because the carrier frequencies of the two QCL combs are unknown, we are, in principle, not able to precisely obtain the frequency link between the microwave and terahertz frequencies. Therefore, we can only give the estimated terahertz frequencies by considering the measured bandwidth and the central frequency of the QCL emission spectra (see Figure \ref{injection}c). To obtain the estimated terahertz frequencies in Figures \ref{filter}b-\ref{filter}d, we assume the central terahertz frequency is 4.2 THz (see Figure \ref{injection}c). For the spectrum at each microwave carrier frequency, the central line is assigned an optical frequency of 4.2 THz. Then, we just need to know the optical bandwidth, which is calculated by $(N-1) f_{\rm{rep}}$ with $N$ being the number of the dual-comb lines and $f_{\rm{rep}}$ the repetition frequency of the QCL combs (here we approximately set $f_{\rm{rep}}$ as 6.15 GHz which is between $f_{\rm{BN1}}$ and $f_{\rm{BN2}}$, see Figure \ref{Schematic}b). The same method is applied to Figure \ref{etalon}a to obtain the estimated terahertz frequencies. After the top axis of Figure \ref{etalon}a is determined, we can add the calculated terahertz transmission of the GaAs etalon in Figure \ref{etalon}a to fit it with the measured dual-comb spectrum.

\medskip
\noindent
\textbf{Supporting Information}\\  
Supporting Information is available from the author.

\medskip
\noindent
\textbf{Acknowledgements}\\ 

This work is supported by the National Natural Science Foundation of China (61875220, 62035005, 61927813, 61991430, and 62105351), the ``From 0 to 1" Innovation Program of the Chinese Academy of Sciences (ZDBS-LY-JSC009), the Scientific Instrument and Equipment Development Project of the Chinese Academy of Sciences (YJKYYQ20200032), the Science and Technology Commission of Shanghai Municipality (21ZR1474600), the National Science Fund for Excellent Young Scholars (62022084), Shanghai Outstanding Academic Leaders Plan (20XD1424700), and Shanghai Youth Top Talent Support Program.
\\

\medskip
\noindent
\textbf{Conflict of Interest}\\ 
The authors declare no conflict of interest.
\\

\medskip
\noindent
\textbf{Data Availability Statement}\\ 
The data that support the findings of this study are available from the corresponding author upon reasonable request.
\\

\medskip

%

\bibliographystyle{MSP}

\bibliography{REF}




\end{document}